\documentclass[twocolumn,
aps,
prl,
amsmath,
amssymb,
floatfix,
reprint]{revtex4-1}

\usepackage{graphicx,xcolor,mathtools,bm,MnSymbol}
\usepackage[normalem]{ulem}
\usepackage{soul}    
\usepackage{verbatim}
\usepackage{verbatim}
\usepackage{esint}

\def \v {{\bf v}}

\def \beq {\begin{eqnarray}}
\def \eeq {\end{eqnarray}}

\def \cb {\color{blue}}

\definecolor{DarkRed}{RGB}{100,0,0}

\definecolor{DarkGreen}{RGB}{0,100,0}

\usepackage[bookmarks=false,linkcolor=blue,urlcolor=blue,colorlinks,citecolor=blue]{hyperref}
\usepackage[all]{hypcap}

\usepackage[english]{babel}
\usepackage[babel,kerning=true,spacing=true]{microtype}

\begin{document}

\title{Specific Heat of a Quantum Critical Metal}

\author{Ori Grossman}
\email{ori.grossman@weizmann.ac.il}
\author{Johannes S.~Hofmann}
\author{Tobias Holder}
\author{Erez Berg}
\email{erez.berg@weizmann.ac.il}
\affiliation{Department of Condensed Matter Physics, Weizmann Institute of Science, Rehovot, 76100, Israel.}

\begin{abstract}
We investigate the specific heat, $c$, near an Ising nematic quantum critical point (QCP), using sign problem-free quantum Monte Carlo simulations. 
Cooling towards the QCP, we find a broad regime of temperature where $c/T$ is close to the value expected from the non-interacting band structure, even for a moderately large coupling strength. At lower temperature, we observe a rapid rise of $c/T$, followed by a drop to zero as the system becomes superconducting. 
The spin susceptibility begins to drop at roughly the same temperature where the enhancement of $c/T$ onsets, most likely due to the opening of a gap associated with superconducting fluctuations.
These findings suggest that superconductivity and non-Fermi liquid behavior (manifested in an enhancement of the effective mass) onset at comparable energy scales. We support these conclusions with an analytical perturbative calculation.
\end{abstract}

\maketitle

\paragraph{ Introduction.---} 
Understanding the continuous formation of order in a Fermi liquid remains a central challenge in the study of strongly correlated electron systems. This problem is complicated by the presence of the gapless quasiparticles at the Fermi surface, which makes the canonical Landau-Ginzburg-Wilson approach of quantum criticality inapplicable~\cite{Hertz1976,Moriya1985,Millis1993,Belitz2005,Loehneysen2007,Nayak1994,Varma2002,Senthil2008}. 
Pinning down the nature of metallic quantum critical points (QCPs) in the case of two spatial dimensions (which is relevant to many quantum materials) has proven particularly challenging ~\cite{Polchinski1994,Kim1994,Oganesyan2001,Metzner2003,Abanov2003,Abanov2004,DellAnna2006,Lee2009,Metlitski2010,Kim2008,Mross2010,Dalidovich2013,Fitzpatrick2013,Metlitski2015,Holder2015,Varma2015,Lunts2017,a_Note}.

The metallic state in the vicinity of the QCP may follow one of two distinct scenarios~\cite{Abanov2001coherent,Metlitski2015,Lederer2015}. 
In the first scenario, a non-Fermi liquid (NFL) metal emerges near the QCP, where electronic quasiparticles become strongly incoherent due to their strong scattering off the critical fluctuations of the order parameter. In the second scenario, superconductivity mediated by the same critical fluctuations gaps the Fermi surface before the non-Fermi liquid develops, and as a result the normal state in the quantum critical regime is a Fermi liquid-like state with coherent quasiparticles.
A successful description of the critical metal has to take into consideration these two competing effects on the same footing. The resulting multi-channel strong coupling problem is currently not amenable to a controlled analytical approach.

A salient feature expected for a nearly-critical Fermi liquid, along with the enhancement of quasiparticle scattering, is a divergence of the effective mass, $m^*$~\cite{Millis1993}. Such a divergence can be probed either by measuring the specific heat coefficient at low temperatures, $c/T\propto m^*$~\cite{Lohneysen1996,Rost2011,Moir2019,Michon2019}, or by other means, such as by measurements of de Haas-van Alphen oscillations~\cite{Walmsley2013,Ramshaw2015}. For example, in the case where the quantum critical fluctuations carry near-zero momentum, a power law divergence of the specific heat is expected; within the random phase approximation (RPA), $c/T\sim T^{-1/3}$~\cite{Halperin1993}. 
However, the divergence of $c/T$ may be preempted by a transition to a superconductor. The main question addressed in this work is whether the quantum critical regime above the superconducting $T_c$ is characterized by a pronounced enhancement of $c/T$ upon approaching the QCP. 

In recent years, it has been demonstrated that models for quantum critical metals can be efficiently simulated using the numerically exact determinant Quantum Monte Carlo (DQMC) method without suffering from the notorious fermion sign problem~\cite{Berg2012,Schattner2016a,Schattner2016,Li2016,Gerlach2017,Wang2017,Xu2017,Li2017,Gazit2017,Wang2018,Liu2019,Bauer2020,Berg2019}. 
Here, we report the first DQMC results for the specific heat of a metallic Ising-nematic QCP at which a discrete $C_4$ rotation symmetry is spontaneously broken. 

Previous works on the this model
focused on the self-energy and transport properties, 
finding signatures of the breakdown of Fermi liquid behavior, 
in an extended temperature window above the superconducting critical temperature $T_c$ near the QCP~\cite{Lederer2017}.
Interestingly, we find that the specific heat in the same temperature regime is close to the non-interacting value down to a temperature $\sim 2 T_c$, 
which we identify as the onset of
superconducting fluctuations, probed by the opening of a spin gap. Thus, the specific heat does not exhibit a broad non-Fermi liquid regime near the QCP.
Inspired by the perturbative structure of the theory, we propose a resolution to this apparent discrepancy between the transport and thermodynamic properties. 

\begin{figure*}[t]
        	\centering\includegraphics[ width=.85\textwidth
        	]{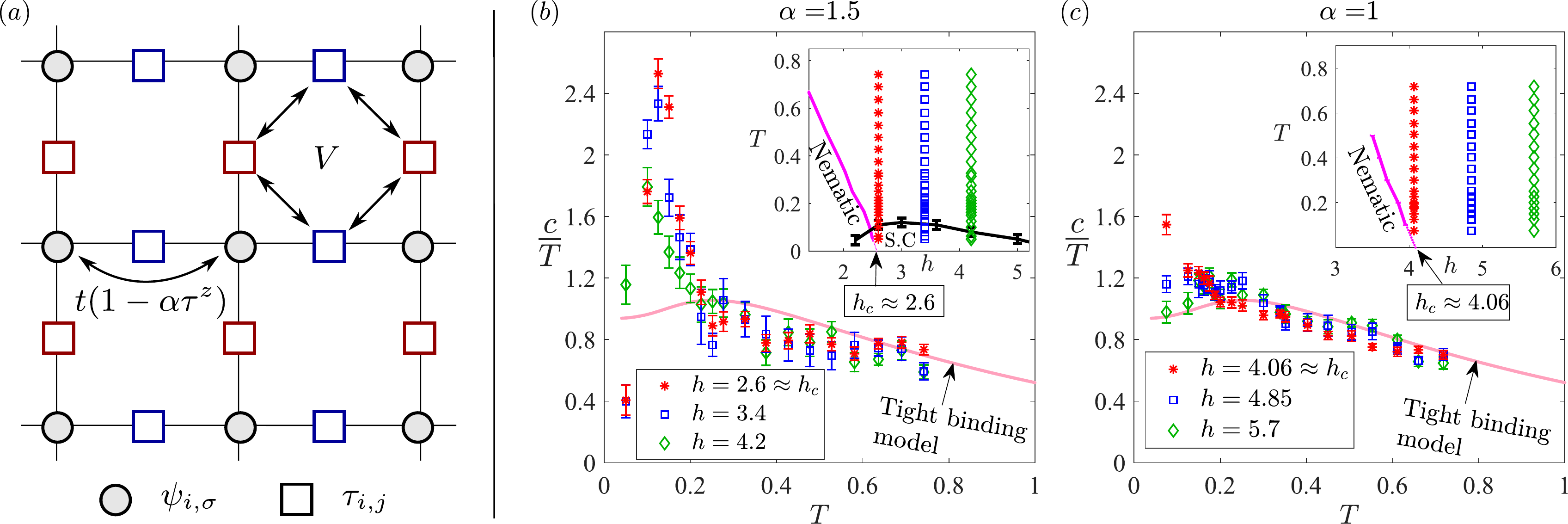}
	\caption{
	 (a) Visualization of the model Hamiltonian.
	 (b,c) $c/T$ as a function of temperature at various $h$ values, for different coupling constants: (b) $\alpha=1.5$, $V=0.5$, and (c) $\alpha=1$, $V=1$. In both cases, $\mu=-1$ and $L=12$. The solid line is $c/T$ of the non-interacting tight binding model. The inset shows the phase diagram in the $(h,T)$ plane. For $\alpha=1.5$, the black line shows the superconducting $T_c$ vs. $h$. For $\alpha=1$, we estimate $T_c\lesssim \ 0.02$ at $h\approx h_c$.}
	 \label{fig:Cv_both_alpha}

\end{figure*} 

\paragraph{Model \& method.---} 
\label{sec:model} 
The model is defined on a
two-dimensional square lattice with a single (spinful) fermionic state per lattice site and a pseudo-spin 1/2 boson on each nearest-neighbor bond (Fig.~\ref{fig:Cv_both_alpha}a)~\cite{Schattner2016}. 
The system is described by the Hamiltonian $H=H_f+H_b+H_{int}$ with
\begin{align}
H_f&= -t
\!\!\sum_{\left<i,j\right>,\sigma}\!\!
\psi_{i,\sigma}^{\dagger}\psi_{j,\sigma} -\mu\sum_{i,\sigma}\psi_{i,\sigma}^{\dagger}\psi_{i,\sigma} \notag\\
H_b&= V
\!\!\sum_{\left<\left<i,j\right>;\left<k,l\right>\right>}\!\!
\tau_{i,j}^{z}\tau_{k,l}^z
-h\sum_{\left<i,j\right>}\tau_{i,j}^{x} 
\notag\\ 
H_{int}&=  \alpha t
\!\!\sum_{\left<i,j\right>,\sigma}\!\!
\tau_{i,j}^{z}\psi_{i,\sigma}^{\dagger}\psi_{j,\sigma} \,.
\label{eq:model_DQMC}
\end{align}
Here, $\psi^{\dagger}_{i,\sigma}$ creates a fermion on site $i$ of spin $\sigma=\downarrow,\uparrow$, $\left<i,j\right>$ denote nearest-neighbour bonds, and $t$, $\mu$ are the hopping amplitude and chemical potential, respectively. The pseudo spins, represented by the Pauli matrices $\tau^{\alpha=x,y,z}_{i,j}$, are governed by a transverse field Ising model. The interaction strength between spins on nearest-neighbor bonds $\left<\left<i,j\right>;\left<k,l\right>\right>$ is given by $V>0$, $h$ sets the transverse field strength, and
$\alpha$ is the dimensionless coupling strength between the pseudospins and the fermions.
Physically, the pseudospins can originate from a purely electronic interactions via Hubbard-Stratonovich transformation, or from bosonic degrees of freedom such as phonons. More importantly, the model is designed to host an Ising nematic critical point, which separates an ordered phase, where the $C_4$ rotational symmetry of the lattice is spontaneously broken, from a $C_4$ symmetric phase. In the ordered phase, the expectation value of $\tau^z$ on horizontal bonds becomes different from that of $\tau^z$ on vertical bonds, breaking the $90^\circ$ rotational symmetry of the lattice. 
The transition can be tuned by the transverse field $h$, and remains continuous down to low temperature~\cite{Schattner2016,Lederer2017}. 

We use the ALF package~\cite{alf_v1}, a general implementation of the auxiliary field quantum Monte Carlo algorithm~\cite{blankenbecler81,assaad08_rev}, to solve the Hamiltonian from above. The negative-sign problem is absent due to time-reversal symmetry for each space-time configuration of $\tau_{i,j}^z$. 
Global updates of the boson fields, that are constructed according to the Wolf algorithm~\cite{Wolff1989}, are used to shorten both the auto-correlation and thermalization times. An artificial orbital magnetic field that couples oppositely to spin up and spin down electrons, corresponding to one flux quantum in the entire system, is applied to reduce finite size effects~\cite{Assaad2002,Schattner2016}. For more details of the QMC implementation, see Refs.~\cite{supp,Schattner2016}. The specific heat $c$ may be evaluated from (i) the numerical derivative of the energy, $c = d\langle H\rangle/dT$, or (ii) the fluctuations of the energy, $c=\beta^2(\langle H^2\rangle - \langle H\rangle^2)$, where $\beta=1/T$~\cite{Notecv}.
In our model, we found that approach (i) converges much faster than (ii)~\cite{supp}.

In the following, we focus on two parameter sets, $(\alpha=1.5,V/t=0.5)$ and $(\alpha=1,V/t=1)$.  
The chemical potential is fixed to $\mu/t=-1$. We point out that there is a van Hove singularity in the band dispersion at $\mu=0$.  
In the Supplementary Material~\cite{supp} we present results for $\alpha=1,\mu/t=-0.5$, where effects of the proximity to the van Hove singularity are more pronounced. We use $t$ as the unit of energy in the remainder.

\begin{figure}[t]
        	\centering\includegraphics[ width=1\columnwidth]{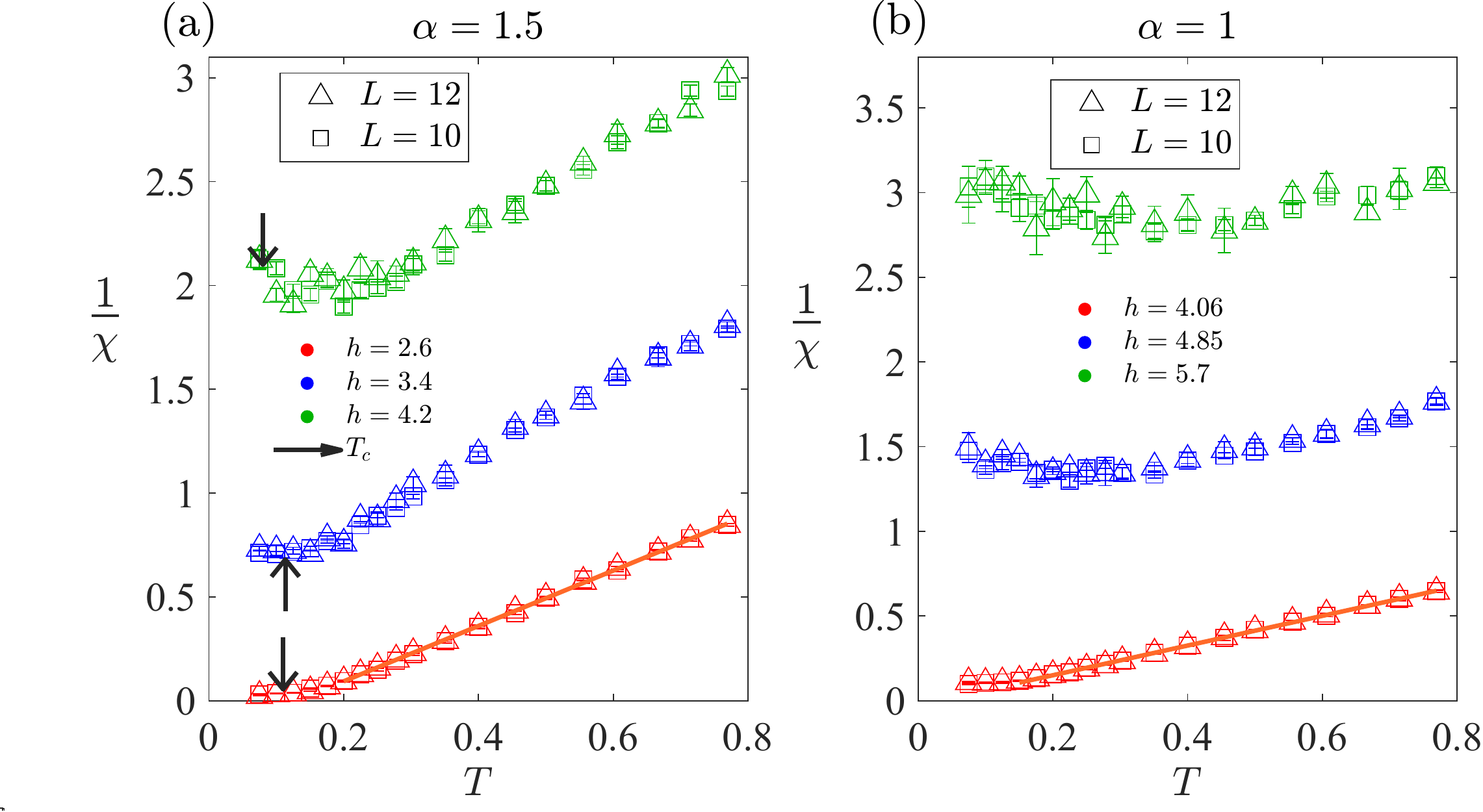}
	\caption{The inverse nematic susceptibility ($1/\chi$) as  a function of temperature,  for two different system sizes ($L=10,12$). In \textbf{(a)} for $\alpha=1.5$ and in \textbf{(b)} for $\alpha=1$. The linear behaviour of $1/\chi$  as a function of $T$ (orange), at $h \approx h_c$, is extended to lower temperatures for $\alpha=1$ . This can be a consequence of the lower $T_c$ at $\alpha=1$, and therefore the fermions are gapped out at lower temperatures. The arrows (black) indicate the superconducting $T_c$ obtained from~\cite{Lederer2017} for $\alpha=1.5$. }
\label{fig:Inv_chi}
\end{figure}

\paragraph{ Results.---} 
We begin by reviewing the phase diagram for the model of Eq.~\eqref{eq:model_DQMC}, described in Refs.~\cite{Schattner2016,Lederer2017}. 
To locate the nematic phase transition, we examine the nematic susceptibility,
\begin{equation}
\chi (h,T)  =
\frac{1}{L^2} \sum_{i,j} \int_0^\beta d\tau \langle N_{i}(\tau) N_{j}(0) \rangle\,,
\label{eq:chi1}
\end{equation}
with the nematic order parameter $N_i = \sum_j \zeta_{ij} \tau^z_{ij}$,
where  $\zeta_{ij}=1/4$ for  $\mathbf{r}_{ij} = \pm \hat{\mathbf{x}}$ (blue squares in Fig.~\ref{fig:Cv_both_alpha}{\cb{a}}), $\zeta_{ij}=-1/4$ for $\mathbf{r}_{ij} = \pm \hat{\mathbf{y}}$ (red squares in Fig.~\ref{fig:Cv_both_alpha}{\cb{a}}), and $\zeta_{ij}=0$ otherwise. $L$ is the linear system size. We present the inverse susceptibility as a function of temperature in Fig.~\ref{fig:Inv_chi} for three transverse field values $h$ and two coupling strengths $\alpha$. The nematic fluctuations are enhanced as the temperature is reduced. $\chi^{-1}$ saturates for the larger values of $h$, which indicates a nematic-disordered ground state, while the susceptibility nearly diverges ($\chi^{-1}\to0$) for the lowest transverse field strength signalling a nematically ordered phase.
The critical transverse field $h_c(T)$ at a given temperature $T$ is determined by a finite size scaling analysis, assuming classical 2D Ising critical exponents~\cite{supp}, and the resulting phase diagram is shown in the inset of Fig.~\ref{fig:Cv_both_alpha} for the two different values of $\alpha$. The quantum critical point is located at $h_c = \lim_{T\rightarrow0} h_c(T)$. 
The superconducting transition temperature $T_c$, extracted from a scaling analysis of the 
s-wave pairing susceptibility~\cite{supp}, also appears in the insets of Fig.~\ref{fig:Cv_both_alpha} for $\alpha=1.5$. For $\alpha=1$, the maximal $T_c$ is smaller than 0.025~\cite{supp} and is not shown.   

We now turn to the specific heat $c(T)$ at and away from the QCP.  Fig.~\ref{fig:Cv_both_alpha}{\cb{b,c}} shows $c/T$ for $\alpha=1.5,\,1$ \ respectively.
At high temperatures, $c/T$ is close to the value computed from the non-interacting tight-binding model (solid line)~\footnote{In this temperature regime, the direct contribution of the bosonic degrees of freedom appears to be small, presumably since these are gapless only close to $q=0$ (as opposed to the fermionic degrees of freedom which are gapless over the entire Fermi surface.)}. 
The broad maximum in the tight binding curve at $T\approx 0.25$  is due to the van Hove singularity in the band structure.
In the stronger coupling case ($\alpha=1.5$), a pronounced peak appears in $c/T$ at low temperatures. 
At $h = 2.6 \approx h_c$, the peak position occurs at $ T_{\mathrm{peak}} = 0.12\pm 0.02$, which is slightly above the superconducting transition at $T_c = 0.1\pm 0.02$. 
We note that, since the superconducting transition is of the Berezinskii-Kosterlitz-Thouless type, the singularity of $c$ at $T_c$ should be very weak, and hence $T_{\mathrm{peak}}$ is not generally expected to coincide with $T_c$. Upon increasing $h$, the peak shifts to a lower temperatures.

At the weaker coupling strength ($\alpha=1$), such a clear peak is absent. 
At the lowest temperatures, there is an enhancement of the specific heat relative to the tight-binding model. This enhancement is most pronounced near the QCP. 
{ It is likely that $c/T$ drops to zero at even lower temperatures, resulting in a finite-$T$ peak in $c/T$, as in the $\alpha=1.5$ case.}

An enhancement in $c/T$ may originate either from an opening of a gap in the quasiparticle spectrum or from an increase of the quasiparticle effective mass~\footnote{We note that, strictly speaking, there is no notion of a gap at finite $T$, due to the presence of thermal excitations.}. 
Evidently, for $\alpha=1.5$, the peak in $c/T$ (Fig.~\ref{fig:Cv_both_alpha}{\cb{b}}) appears both near the QCP and away from it, hence it likely not caused by an enhanced $m^*$ due to quantum critical fluctuations. The shift of the peak position upon increasing $h$ mirrors the decrease of the superconducting critical temperature, suggesting that the superconducting gap opening is the main source of the peak in $c/T$.
The situation is less clear for the weaker coupling $\alpha=1$ (Fig.~\ref{fig:Cv_both_alpha}{\cb{c}}), where a significant enhancement of $c/T$ is only detectable near $h=h_c$.

\begin{figure}[t]
        	\centering\includegraphics[ width=1\columnwidth,  ]{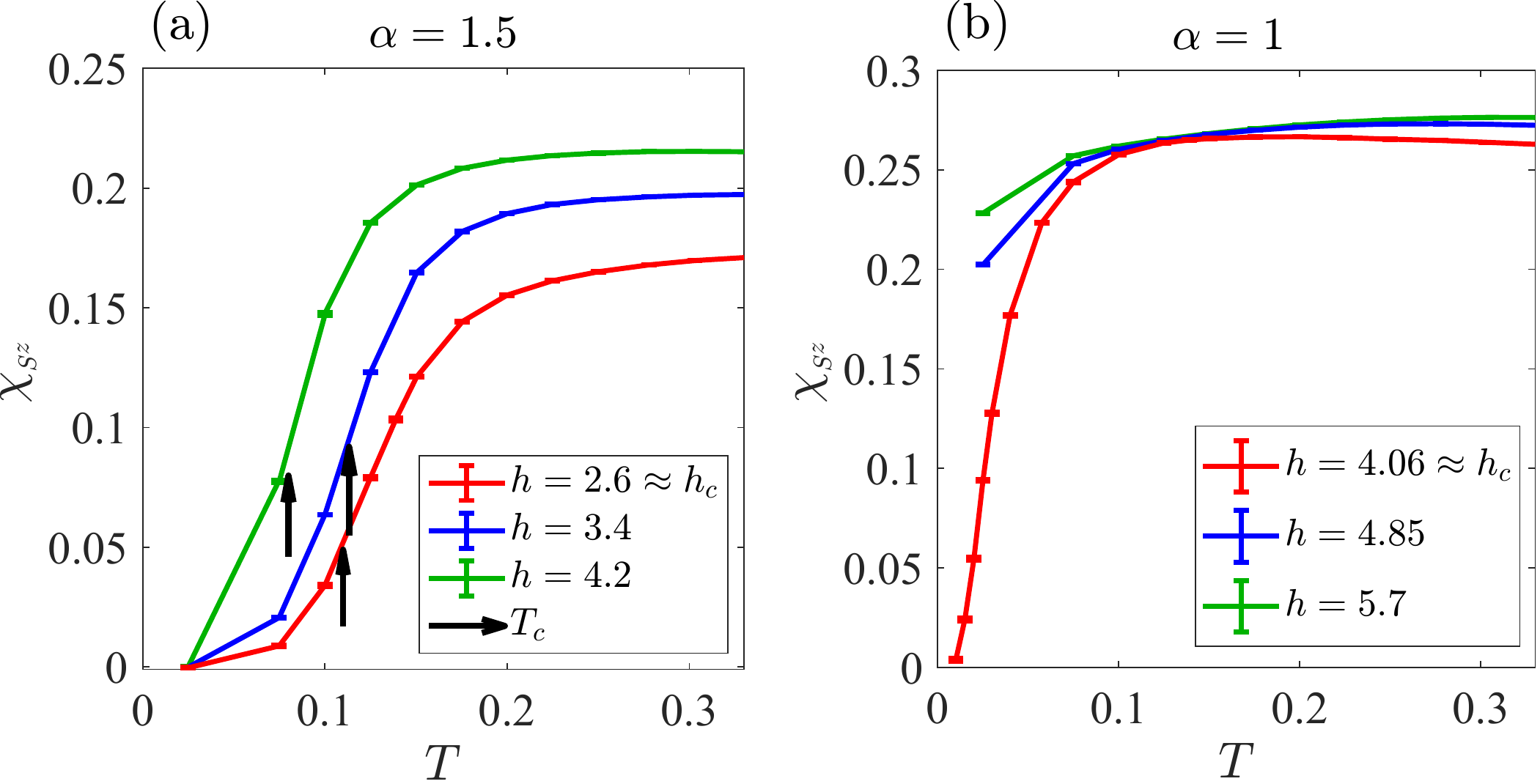}
	\caption{The spin susceptibility $\chi_{{S^z}}$ as a function of temperature at various $h$ values for the two different coupling constants.
	In \textbf{(a)} for $\alpha=1.5$ and in \textbf{(b)} for $\alpha=1$. As before, The arrows (black) indicates the superconducting $T_c$ obtained from~\cite{Lederer2017} for $\alpha=1.5$. The system size is $L=12$.}
		\label{fig:Spin_sus}
\end{figure}

In order to identify the origin of the low-temperature enhancement of the specific heat, we study the spin susceptibility,
\begin{equation}
\chi_{S^z}=\frac{1}{L^2}\sum_{i,j} \int_{0}^{\beta} \ d\tau \left<\hat{S}^z_{i}(\tau)\hat{S}^z_{j}(0) \right> \,,
\label{eq:chi_Sz}
\end{equation}
where  $\hat{S}^z_{i} = \frac{1}{2} \left( \psi^{\dagger}_{i,\uparrow} \psi_ {i,\uparrow} -
\psi^{\dagger}_{i,\downarrow}  \psi_ {i,\downarrow} \right)$.
$\chi_{S^z}$ is shown in Fig.~\ref{fig:Spin_sus}. 
The spin susceptibility is roughly constant for $T > 0.2$ ($T>0.1$) for $\alpha=1.5$ ($\alpha=1$).
As the temperature is lowered further, $\chi_{S^z}$ begins dropping dramatically, consistent with the opening of a spin gap. 

The temperature where the suppression of $\chi_{S^z}$ onsets is comparable to the temperature at which $c/T$ begins to rise (Fig.~\ref{fig:Cv_both_alpha}). This leads to an interpretation of both the enhancement of $c/T$ and the suppression of $\chi_{S^z}$ as a signature of an opening of a gap, most likely associated with superconducting fluctuations. Further evidence for this interpretation is provided by a  rapid growth of the superconducting susceptibility and a suppression of the single-particle density of states near the Fermi level, which both onset at a similar temperature~\cite{supp}. Note that this implies that the gap at $h\approx h_c$ onset at a temperature significantly larger than the superconducting critical temperature, $T_c\approx 0.1$ for $\alpha=1.5$ (marked by black arrows in Fig.~\ref{fig:Spin_sus}{\cb{a}}), and $T_c\lesssim 0.02$ for $\alpha=1$~\cite{supp}. Such a regime is commonly referred to as a ``pseudogap regime'' (or a regime of ``preformed Cooper pairs'' without long-range phase coherence). This behavior is in contrast to the expectation from weak-coupling mean-field theory, which predicts a gap that onsets concomitantly with $T_c$, but is in agreement with prior QMC results in models of quantum critical metals at intermediate to strong coupling~\cite{Schattner2016a,Lederer2017}. 

The temperature range where $\chi_{S^z}$ is approximately constant, above the onset of the spin gap, can be interpreted in terms of a Fermi liquid. This interpretation is supported by the fact that, in the same regime, $c/T$ is not strongly temperature dependent, as shown in Fig.~\ref{fig:Cv_both_alpha} (and its temperature dependence can mostly be ascribed to band structure effects). In addition, the single-particle density of states at the Fermi level is found to be weakly temperature dependent in the same regime~\cite{supp}. Within Fermi liquid theory, $c/T \propto m^*$, the single-particle density of states is proportional to $Z m^*$ where $Z$ is the quasiparticle weight~\cite{b_Note}, and $\chi_{S^z}= \pi^{-2} (1+F^a_0)^{-1} k_F m^*$. Here, $F^a_0$ is the isotropic, spin-antisymmetric component of the Landau quasiparticle interaction. Hence, at temperature above the onset of a spin gap, we can explain our data qualitatively by assuming that for $\alpha=1$, $Z\approx 1$ and $F_0^a$ is small. For $\alpha=1.5$, in contrast, $Z$ decreases substantially and $F_0^a$ grows as $h$ approaches $h_c$.  

Combining the above arguments suggests the following picture for the behavior at $h\approx h_c$: 
(i) In a broad temperature range below the Fermi energy $E_F$, the system's thermodynamic properties are roughly consistent with Landau's Fermi liquid theory. (ii) Below a certain temperature, smaller than $E_F$ but significantly larger than the superconducting $T_c$, $c/T$ is enhanced, more or less concomitantly with a suppression of the spin susceptibility and the single-particle density of states. All these effects are most probably due to the onset of a gap due to superconducting fluctuations. (iii) At the lowest temperatures (below $T_c$) superconductivity is established.

\paragraph{ Perturbation theory.---}
\label{sec:RPA}
It is useful to relate our findings at strong coupling with the results of the standard RPA analysis~\cite{Halperin1993}.
The perturbative calculation is controlled in the limit of a large number $N$ of fermion species (the physical value is $N=2$) and not too low temperatures, as discussed below. 
For simplicity, we consider a system with dispersion $\varepsilon(\bm{k})=\frac{\bm{k}^2}{2m}-\mu$. The four-fermion interaction is taken to be of the form 
\begin{align}
U(\bm{q},\bm{k},\bm{k'})=\frac{g^2f_{\bm{k}-\bm{q}/2}f_{\bm{k'}+\bm{q}/2}}{N(r_0+|\bm{q}|^{2})}
\end{align}
where $\bm{q}$ is the momentum transfer, $g^2$ is the coupling strength, and $f_{\bm{k}}$ is the nematic form factor: $f_{\bm{k}}=\cos k_{x}-\cos k_{y}$. The parameter $r_0$ is used to tune the system to the QCP, which occurs at $r_0 = r_c>0$. 

At $r_0 = r_c$, the specific heat is given by~\cite{supp}
\begin{align}
  c(T)=
  \frac{\pi}{6}mNT \bigg[ 1
  +\frac{A}{N}\left(\frac{g^4}{E_F T}\right)^{\frac{1}{3}} + O\left(\frac{1}{N^2}\right)\bigg],
  \label{eq:RPA}
\end{align}
where $A\approx 0.19$. 
The second term in the square brackets in Eq.~\eqref{eq:RPA} describes the enhancement of the specific heat due to quantum critical fluctuations. This term becomes significant compared to the first (non-interacting) term at a temperature $T_{\mathrm{NFL}} \propto
\frac{g^4}{N^3 E_F}$. $T_{\mathrm{NFL}}$ is the temperature below which electrons lose their coherence, and the Fermi liquid description breaks down. At the same temperature, terms which are naively of higher order in $1/N$ become parametrically enhanced, and the $1/N$ expansion is no longer controlled~\cite{Lee2009,Metlitski2010,Holder2014}. Moreover, solving the linearized Eliashberg equation for the pairing vertex gives that $T_c\propto T_{\mathrm{NFL}}$~\cite{Wu2020,supp}. Thus, the weak coupling analysis predicts no parametric separation in temperature between the breakdown of Fermi liquid theory (manifested as a divergence of $c/T$) and the onset of a pairing gap.
This conclusion is corroborated by an RPA analysis of a lattice model including the full tight-binding dispersion [Eq.~\eqref{eq:model_DQMC}], showing no significant deviation of $c/T$ relative to the non-interacting value for $T\gtrsim 0.2$~\cite{supp}.
These observations mirror the picture that emerges at moderate to strong coupling from our QMC results, i.e., the enhancement of $c/T$, relative to the value expected from the band structure, onsets at the same temperature where a pairing gap appears. 

\paragraph{Discussion.---}
\label{sec:conc}
In this work, we have examined the specific heat in the vicinity of a quantum critical point in a metal, using unbiased, numerically exact QMC simulations. We find that, upon cooling the system towards the quantum critical point, $c/T$ is enhanced relative to the band structure value. The enhancement of $c/T$ onsets at roughly the same temperature where the spin susceptibility and the single-particle density of states exhibit a downturn, signalling
the appearance of a gap, most likely due to the onset of superconducting fluctuations. 
$c/T$ is suppressed sharply upon entering the superconducting phase which covers the QCP. 

Thus, our main conclusion is that within our model, there is no broad non-Fermi liquid regime characterized by a diverging $c/T$ near the QCP. This is most probably because the quantum critical enhancement of $c/T$ is preempted by the opening of a pairing gap. These observations are qualitatively consistent with the expectation from the weak-coupling RPA analysis, which predicts that $T_{\text{NFL}}$ and $T_c$ are of the same order of magnitude. 
However, in our simulations (performed in moderate to strong coupling) we find that the gap due to superconducting fluctuations appears at a temperature significantly above $T_c$, in contrast to the weak-coupling analysis in which the superconducting transition is essentially mean-field like.

Above the gap opening temperature, we find a broad
temperature regime where $c/T$ shows no significant enhancement relative to the band structure value. This is surprising, since in this model, the same temperature regime has been shown to exhibit strong deviations from Fermi liquid theory in the frequency dependence of the self-energy and the temperature dependence of the transport scattering rate~\cite{Lederer2017,Klein2020}. This apparent discrepancy can be understood from the fact that, in this regime, the quasiparticles are still coherent (in the sense that their self-energy $\Sigma(i\omega_n)$ is smaller than $\omega_n$, even at the smallest Matsubara frequency~\cite{supp}), and the effective mass is not significantly enhanced. However, the self-energy (and hence the scattering rate) has a non-Fermi liquid temperature dependence~\cite{DellAnna2007,Maslov2011,Wang2019}. 

We end with two comments regarding the implications of our study for experiments in quantum materials. First, our results imply that observing a divergence of $m^*$ near a QCP generally requires suppressing superconductivity, e.g., by applying a magnetic field (which is unfortunately impossible in our simulations without introducing a fermion sign problem). Second, it is interesting to note that in FeSe$_{1-x}$S$_x$, a broad regime of quasi-linear resistivity is observed near the putative nematic QCP~\cite{Licciardello2019,Huang2020,Bristow2020} with no accompanying discernible enhancement of $m^*$~\cite{Coldea2019}. These findings may be explained by the presence of coherent quasiparticles scattered by quantum critical fluctuations, analogous to the behavior found in our model.

\begin{acknowledgements}
We thank 
A. Chubukov, S. Kivelson, and Y. Schattner 
for useful discussions. 
This work was supported by the European Research
Council (ERC) under grant HQMAT (Grant Agreement
No. 817799), the US-Israel Binational Science Foundation (BSF) under grant no. 2018217, the Minerva foundation, and a research grant from Irving and Cherna Moskowitz. The auxillary field QMC simulations were carried out using the ALF package available at \url{https://alf.physik.uni-wuerzburg.de}.
\end{acknowledgements}

\bibliography{main_arxiv.bbl}

\begin{thebibliography}{75}
\bibitem [{Not()}]{Notecv}%
  \BibitemOpen
  \href@noop {} {}\bibinfo {note} {Note that this is the specific heat at a
  fixed chemical potential, rather than at a fixed density.}\BibitemShut
  {Stop}%
\bibitem [{b_N()}]{b_Note}%
  \BibitemOpen
  \href@noop {} {}\bibinfo {note} {Here, for simplicity, we assume $Z$ and
  $m^*$ to be direction-independent.}\BibitemShut {Stop}%
\end{thebibliography}

\clearpage

\appendix

\onecolumngrid
\begin{center}
\huge{Supplemental Material}
\end{center}

\vspace*{\baselineskip}

In this supplementary material, we elaborate on the techniques we used to measure the specific heat and determine the phase transition lines. Additional data and analysis is presented for the single-particle density of states at the Fermi level. We elucidate in some more detail the role of the van Hove singularity in the band structure. We present data showing the finite size effect on our QMC results. Finally, we present details of the perturbative RPA analysis.

\vspace*{2\baselineskip}

\twocolumngrid

\renewcommand{\thefigure}{S\arabic{figure}}
\renewcommand{\figurename}{Figure}
\setcounter{figure}{0}
\setcounter{equation}{0}
\renewcommand{\theequation}{S.\arabic{equation}}

\section{Technical details regarding the DQMC simulations} 
We are using the same setup of the DQMC simulation here as discussed in Ref.~\cite{Schattner2016}.
The DQMC relies on a Trotter decomposition of the partition sum with the inverse temperature $\beta=N\Delta\tau$. We use $\Delta\tau=0.05$ throughout the manuscript, which introduces a systematic Trotter error of order $\mathcal{O}({\Delta\tau}^2)$. We have confirmed that the results are converged and do not depend of this choice of $\Delta\tau$.

As mentioned in the main text, we also use the Wolf algorithm for global updates in order to reduce the autocorrelation and warmup times. We design a cluster of spins to be updated solely according to the bosonic part $H_b$ of the action. The acceptance of this cluster update is determined according to the contribution of the rest of the Hamiltonian, $H_f+H_{int}$, to the Boltzmann weight (given by the fermion determinant). This procedure respects detailed balance. 
The optimal number of global moves per sweep of single spin flips depends on the system size $L$ and inverse temperature $\beta$, as well as on the distance from the critical point.
Typically, we use a few hundred global updates per single spin-flip sweep.

In addition, we reduced finite size effects by including an orbital 
pseudo-magnetic field that couples oppositely to spin up and spin down electrons, preserving time-reversal symmetry. There is a single flux quantum through the entire system, such that the field vanishes in the thermodynamic limit~\cite{Assaad2002}. 
This breaks translation symmetry, allowing for a level repulsion between single-particle energy levels which leads to a smoother density of states and consequently to a rapid convergence towards the thermodynamic limit.

\section{Measurement of the specific heat} 
\label{app:Cv_methods}
The specific heat $c$ is known to be a challenging quantity to compute in determinantal quantum Mote Carlo simulations. Two equivalent definitions of $c$ can be used:
\begin{align}
	&\quad&  c&=\frac{\beta ^2}{L^2} \left( \langle H^2 \rangle - \langle H \rangle ^2 \right),\label{eq:cVmethod1}
\\
	&\quad& c&=\frac{1}{L^2}\cdot \frac{\partial \langle H\rangle}{\partial T} \,.
	\label{eq:cVmethod2}
\end{align}
In the following we list a few arguments concerning the statistical errors of the two approaches.

\begin{figure}
	\centering\includegraphics[ width=1\columnwidth ]{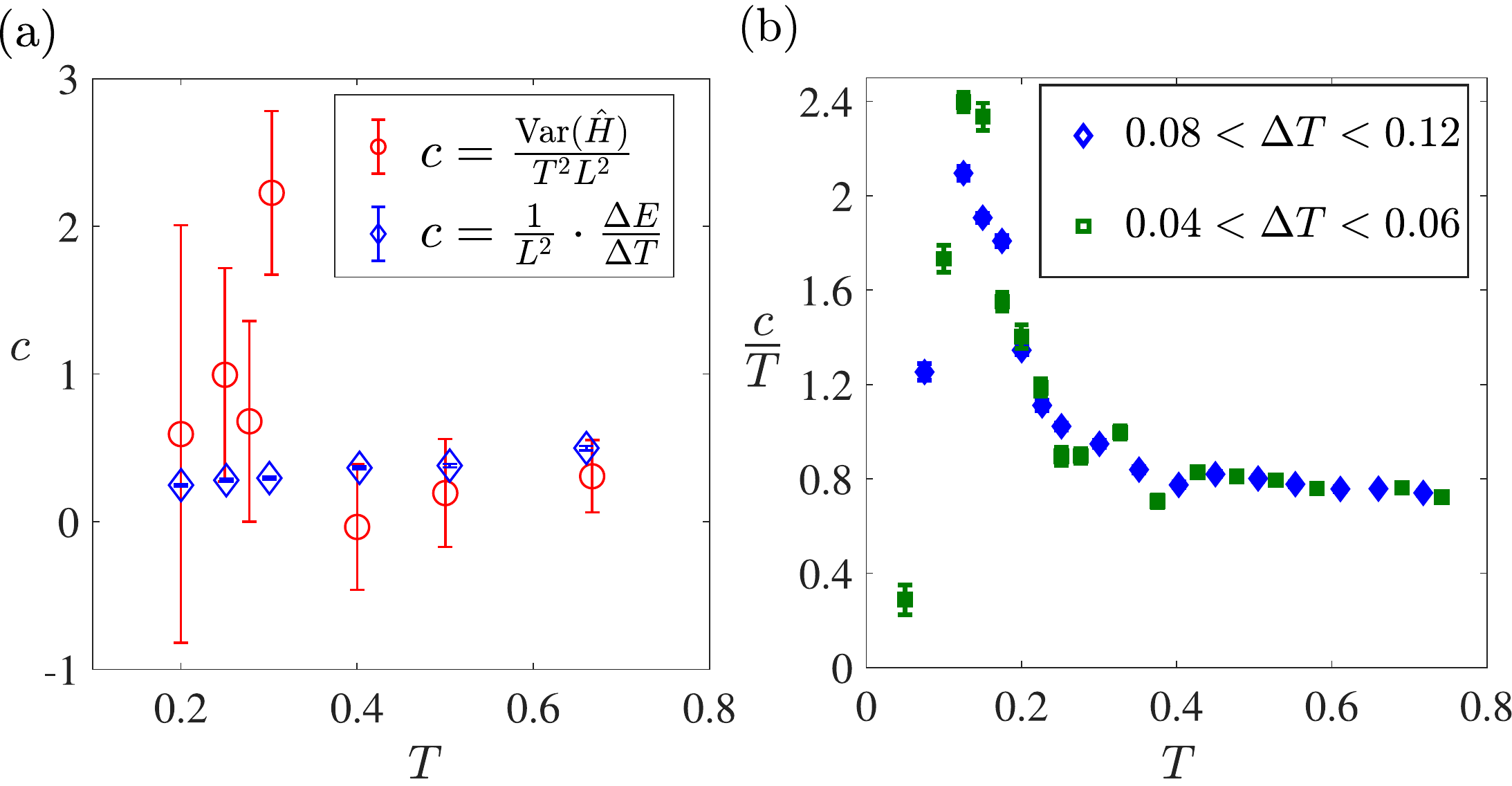}
	\caption{Comparison between the different methods of evaluating the specific heat. \textbf{(a)} shows the calculation by using the energy derivative (Eq.~\eqref{eq:cVmethod1}, blue) and  the  variance of the Hamiltonian (Eq.~\eqref{eq:cVmethod2}, red). Using a numerical derivative is significantly better in terms of the statistical errors.  \textbf{(b)} shows the estimate of $c$ from the numerical derivative using different values for $\Delta T$ (nearest and next nearest neighbours points of $E\left(T\right)$).}  
\label{fig:cv_method_plot}
\end{figure}

Note that in a Fermi liquid $c \propto T$, such that $\langle H^2 \rangle - \langle H \rangle ^2 \propto T^3$ and $\langle H \rangle = E_0 + \mathrm{const.}\times \,T^2 + \mathcal{O}(T^3)$. 
Hence, the variance of the energy is a difference of two, much larger quantities, unless $E_0=0$. Thus, a Monte Carlo estimation of Eq.~\eqref{eq:cVmethod1} typically suffers from large statistical errors. This problem is partially mitigated by measuring $\langle H^2 \rangle$ and $\langle H \rangle^2$ using the same Monte Carlo configurations, such that their errors tend to cancel. 

Additionally, the calculation of $\langle H^2\rangle$ involves a high order correlation functions, e.g., computing $H_{int}^2$ includes the calculation of $6$-point correlators. Typically, in Monte Carlo, the statistical error increases with the the order of the correlation function. 
The expectation value of an observable $O$ is $\langle O \rangle = \sum_C p(C) \llangle O \rrangle_C $ where $C$ is a given Monte Carlo configuration, $p(C)$ the configuration's probability, and $\llangle O \rrangle_C$ is the value of the observable for the given configuration. It is not guarantied that a configuration with high probability $p(C)$ also has a large contribution to the observable. Instead, there may be configurations with a small $p(C)$ but a large $p(C) \llangle O \rrangle_C$. In such a case, the statistical error can be very large, since the observable is dominated by rare configurations. 
Such a mismatch between the configurations' probability and their contribution to the observable is more likely for high order observables.

The alternative approach, Eq.~\eqref{eq:cVmethod2}, extracts the specific heat from the energy $E(T) = \langle H \rangle$ as a function of temperature via a numerical derivative $\frac{E(T+\Delta T/2)-E(T-\Delta T/2)}{\Delta T}$, where $\Delta T$ should be sufficiently small. Here, one has to distinguish two different sources of error. On one hand, there is the statistical error in $E(T+\Delta T/2)-E(T-\Delta T/2)$, which decreases with increasing $\Delta T$. On the other hand, there is a systematic error in the estimate of the derivative that increases with $\Delta T$. 

\begin{figure}[tb]
\includegraphics[width=1.0\columnwidth]{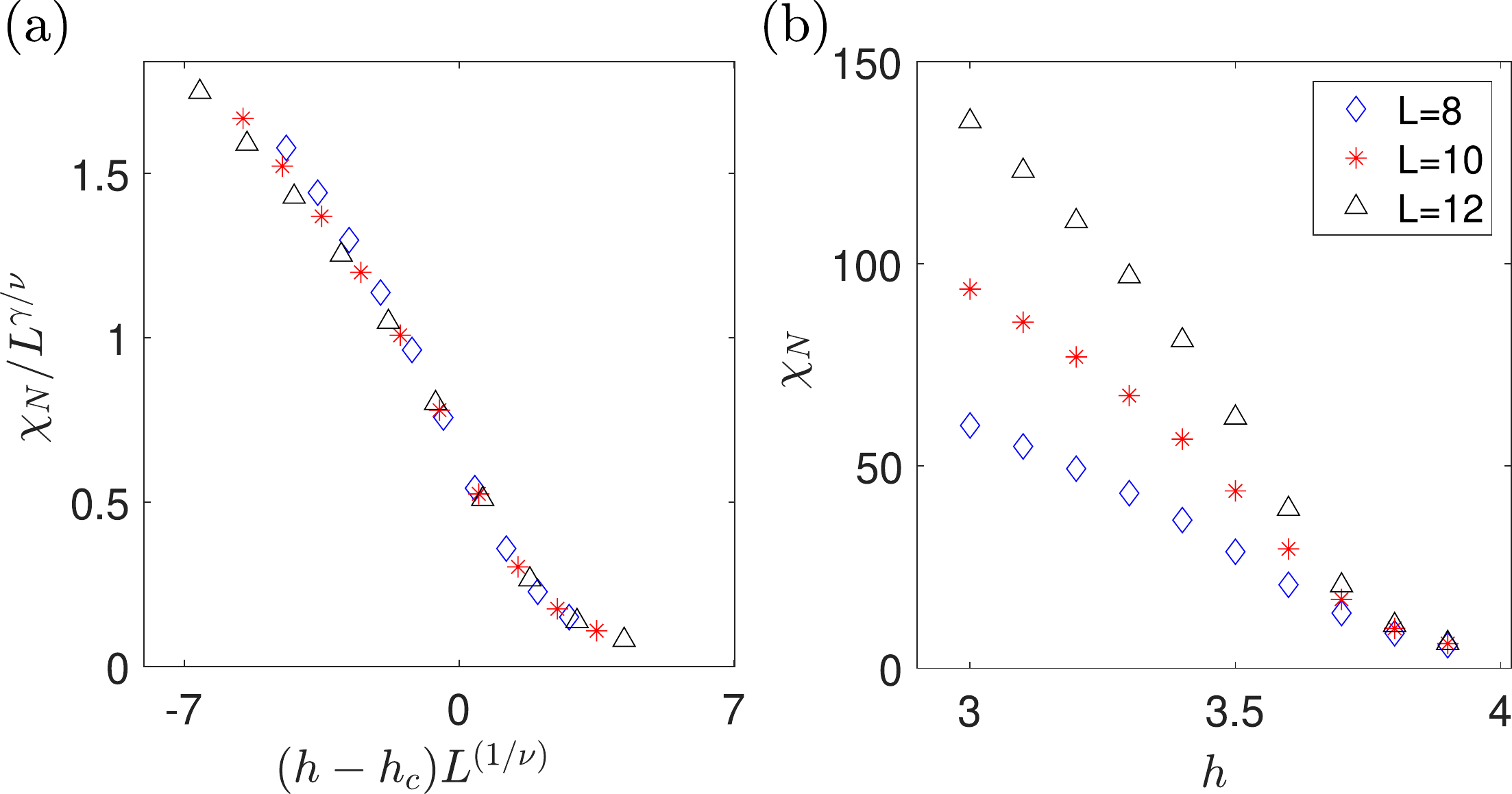}
\caption{An example of the finite size scaling employed to calculate the critical value of $h$ for a given temperature. On the left, the rescaled susceptibility is shown as a function of the rescaled field for three system sizes, which collapse onto a universal function. On the right, the data is shown without rescaling. The parameters are $\alpha=1$, $\mu=-0.5$ and $T=0.5$, with the resulting $h_c\left(T\right) \approx 3.55$.}
\label{fig:Ising2d_collapse}
\end{figure}
\begin{figure}
    \includegraphics[width=1.0\columnwidth]{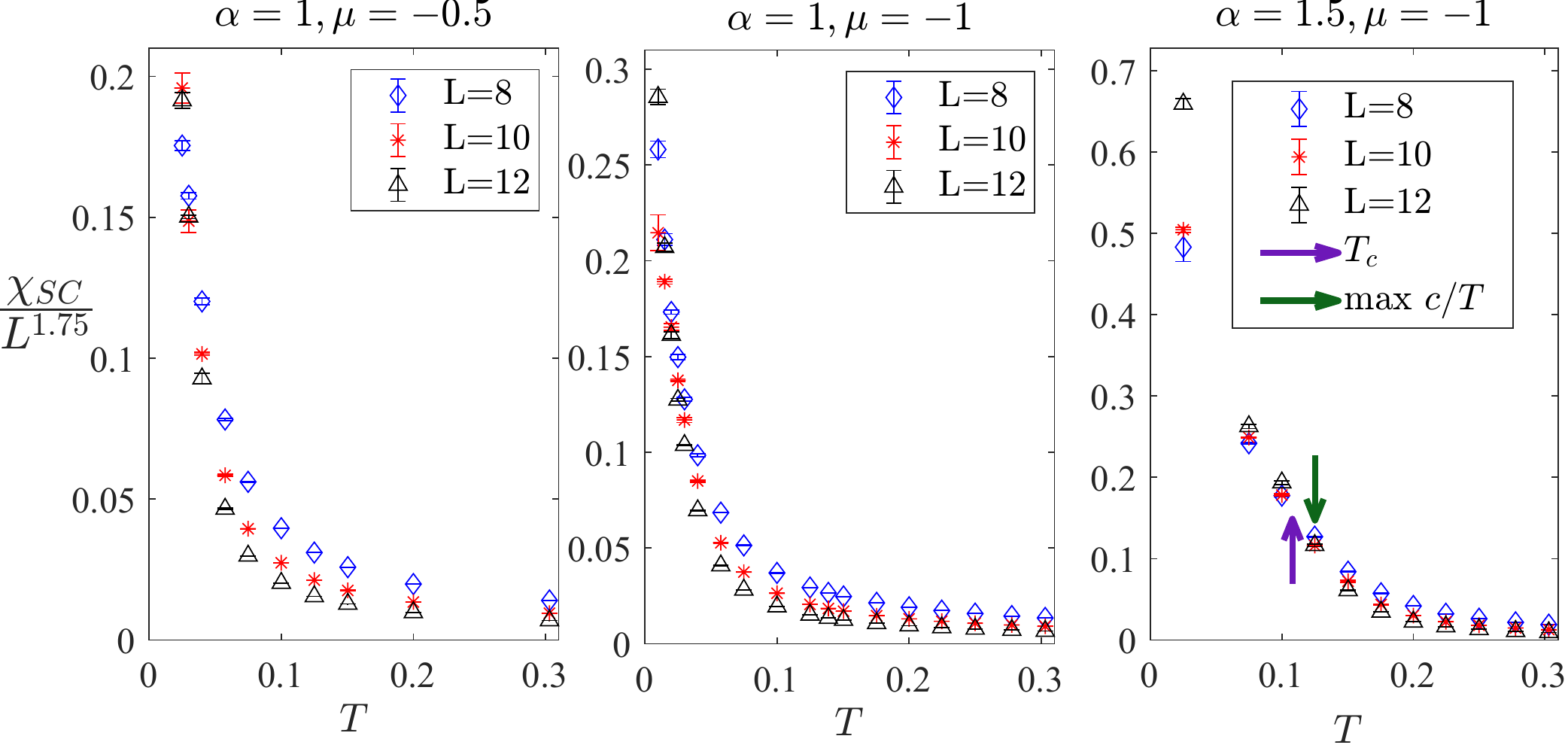}
\caption{ Finite size scaling for the superconducting susceptibility. For $\alpha=1.5$, we see a clear crossing at $T\approx 0.1$ , after which the dependence on system sizes reverses its order, as expected below $T_c$ (Marked by a purple arrow). We also indicate the temperature at which $c/T$ is maximal (green arrow).   For $\alpha=1$, no clear crossing occurs until $T\approx 0.025$ for $\mu=-0.5$ and $T\lesssim 0.02$ for $\mu=-1$.}
\label{fig:sc_sus_scaling}
\end{figure}
\begin{figure}
        	\centering\includegraphics[ width=1\columnwidth ]{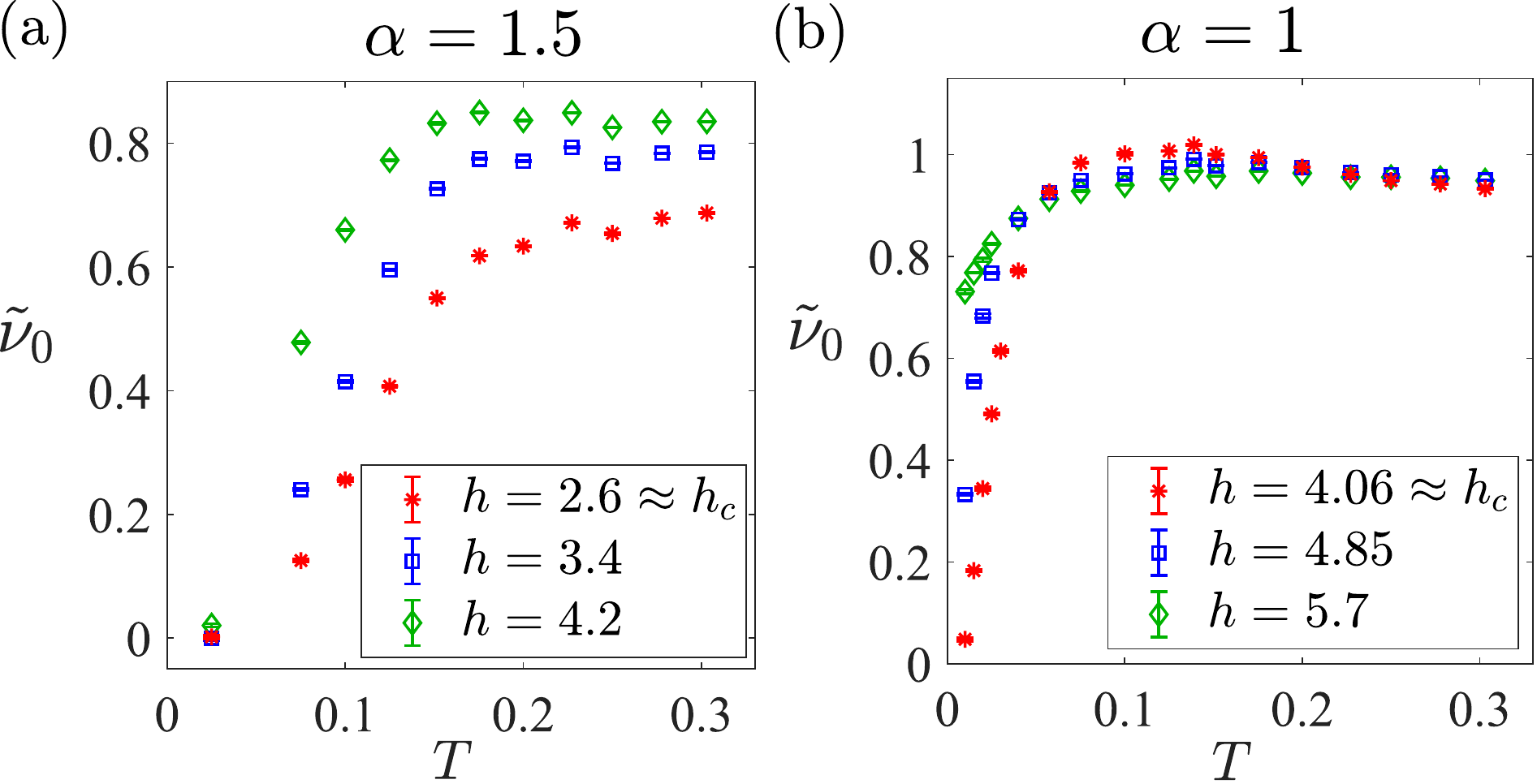}
	\caption{Proxy for the single-particle DOS near the Fermi level, $\tilde{\nu}_0(T)$ (see Eq. \ref{eq:DOS_def}) as a function of temperature for various $h$ values. \textbf{(a)} $\alpha=1.5$, \textbf{(b)} $\alpha=1$. The system size is $L=10$. In both cases, $\tilde{\nu}_0(T)$ starts decreasing at a temperature significantly above the superconducting $T_c$. The onset of the suppression of $\tilde{\nu}_0$ occurs close to (but slightly below) the temperature where the enhancement of $c/T$ begins (see main text).}  
		\label{fig:GF_main}
\end{figure}

A priori, it is not obvious which of the two approaches is more efficient for a fixed amount of computing resources. A comparison is presented in Fig.~\ref{fig:cv_method_plot}{\cb{a}} where we show the specific heat determined via Eq.~\eqref{eq:cVmethod1} and Eq.~\eqref{eq:cVmethod2} using a similar computing time. We observe a much smaller statistical errors for the numerical derivative compared to the variance of the energy. Also note that the statistical uncertainty is increasing at lower temperatures.
In Fig.~\ref{fig:cv_method_plot}{\cb{b}} we show the numerical derivative results for two values of $\Delta T$. The result does not strongly depend on $\Delta T$. In particular, the data for $T>0.3$ seems to have converged to the limit $\Delta T\rightarrow 0$. Also, the position of the peak in $c/T$ does not depend on the choice of $\Delta T$, even though the height of the peak is somewhat reduced for the larger $\Delta T$ values. 

\section{Finite size scaling} 
\label{app:Finite_size_scaling}
The finite-temperature nematic phase boundary can be estimated by using 
finite-size scaling techniques for the classical two-dimensional Ising
transition, characterized by a correlation length critical exponent $\nu=1$ and a susceptibility critical exponent $\gamma=7/4$. In the vicinity of $T_c$, the nematic susceptibility $\chi_{{N}} $  satisfies:
\begin{equation}
 \chi_{{N}}\left(h,L\right)=L^{\gamma/\nu}F\left((h-h_c)L^{1/\nu}\right),
\end{equation}
where $F$ is a universal scaling function. We use different system sizes and estimate by data collapse the critical transverse field for a given temperature. An example of this kind of procedure is illustrated in Fig.~\ref{fig:Ising2d_collapse}.

\begin{figure*}
	\centering\includegraphics[ width=.98\textwidth ]{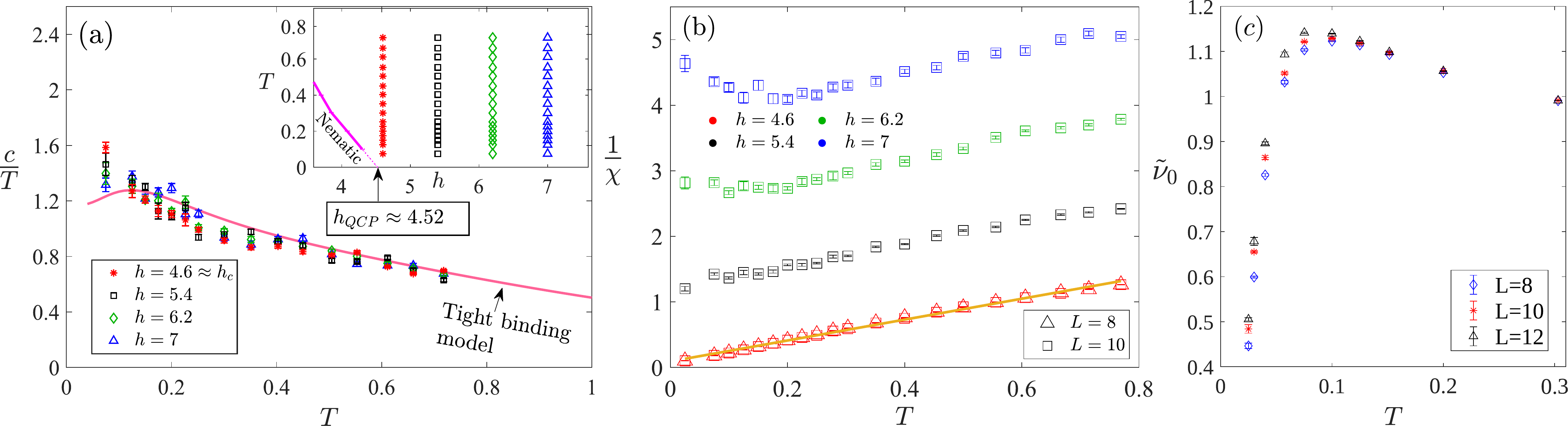}
	\label{fig:CV_a1mu0_5_App}
	\label{fig:Inv_chi_App}
	\caption{Results for $\mu=-0.5$, $V=1$, $\alpha=1$, where the effects of the nearby van Hove singularity in the band structure are more pronounced than in Fig. 1 of the main text. 
	(a) $c/T$ as a function of temperature at various $h$ values, for system size $10\times 10$. The inset shows the phase diagram and the $(h,T)$ values of the points where $c$ was measured. The solid curve shows $c/T$ for the tight-binding model. 
	(b) The inverse nematic susceptibility $1/\chi$ as a function of temperature, for $L=8,10$ and different values of $h$. At $h \approx h_c$, $1/\chi(T)$ is linear in $T$ down to the lowest temperatures (orange line). 
	(c) Proxy for the single-particle DOS at the Fermi level, $\tilde{\nu}_0$ [Eq.~\eqref{eq:DOS_def}]. The suppression of $\tilde{\nu}_0$ onsets significantly above superconducting transition, $T_c\approx 0.025$.  
	} 
	\label{fig:DOS_a1mu0_5_App}
\end{figure*} 
The same finite size scaling method is used to calculate $T_c$ of the superconducting BKT transition.
{We have computed the $s$--wave  superconducting susceptibility $\chi_{SC}$, which is defined as 
\begin{equation}
\chi_{SC} = \frac{1}{L^2}\int_0^\beta \!\mathrm{d}\tau \sum_{i,j} \langle \Delta_{i}{\vphantom{\dagger}} (0)  \Delta^\dagger_{j}(\tau) \rangle
\label{eq:sc_sus}
\end{equation}
where $\Delta_i(\tau) = c_{i\uparrow}(\tau) c_{i\downarrow}(\tau)$.}
In the vicinity of the critical temperature, the finite size scaling of the superconducting  susceptibility satisfies~\cite{Moreo1991}
\begin{equation}
\chi_{{SC}}=L^{2-\eta(T_c)}f(L/\xi),
\end{equation}
where $\eta(T)$ increases monotonically with increasing $T$ between $\eta(T=0)=0$ and
$\eta(T=T_c)=\frac{1}{4}$. We can therefore estimate $T_c$ by locating the crossing point between different system sizes upon setting $\eta(T=T_c)=\frac{1}{4}$.
An example for the estimation of $T_c$ using this procedure is illustrated in Fig \ref{fig:sc_sus_scaling}.

\section{Density of states }
To shed further light on the origin of the low-temperature enhancement of $c/T$, we study the integrated density of states (DOS) around the Fermi level in an energy window set by the temperature. We recall the relation of the fermionic spectral function $A(\mathbf{k},\omega)=-\pi^{-1}\mathrm{Im}\,G_R(\mathbf{k},\omega)$ (where $G_R$ is the retarded Green's function) and the imaginary time-displaced Green's function \cite{Trivedi1995}\looseness=-1
\begin{equation}
G(\mathbf{k},\tau=\beta/2) = \int_{-\infty}^\infty d\omega \frac{ A(\mathbf{k}, \omega)}{2\cosh({\beta \omega }/2)} .
\label{eq:G3}
\end{equation}
We further introduce the following integral over the Green's function 
\begin{equation}
\tilde{\nu}_0 \triangleq \beta\int d^2 k \ G(\mathbf{k},\tau=\frac{\beta}{2}),
\label{eq:DOS_def}
\end{equation}
which can be considered a proxy for the single-particle DOS at the Fermi level.
Then, an opening of the gap is manifested by a rapid suppression of $\tilde{\nu}_0$ at low temperatures (Fig.~\ref{fig:GF_main}). We observe that for $\alpha=1$, $\tilde{\nu}_0$ undergoes a downturn at $T = T_{fluc} \approx 0.08$, which we interpret as an onset of a gap due to superconducting fluctuations. $T_{fluc}$ is comparable to the temperature at which the enhancement of $c/T$ onsets, suggesting that the enhancement may at least partially due to superconductivity.
Although for $\alpha=1$ we could not precisely estimate $T_c$, the divergence of the superconducting susceptibility (Fig.~\ref{fig:sc_sus_scaling}) strongly suggests that the gap opening is also due to superconducting fluctuations. Within our resolution it seems that $T_c\lesssim \ 0.02$ (see Fig.~\ref{fig:sc_sus_scaling}).  
The superconducting fluctuations appear at a surprisingly high temperature, $T_{fluc} \approx 4T_c$. This broad fluctuation-dominated regime is qualitatively distinct from the expectation according to mean-field BCS theory.

\section{  Approaching the van Hove singularity } 
\label{app:more_data_alpha1_mu0_5}
To test the effect of the proximity to the van Hove singularity, we have studied the case of moderate coupling $\alpha=1$ with $\mu=-0.5$ (which brings us closer to the van Hove singularity relative to the case $\mu=-1$, studied in the main text). Fig.~\ref{fig:CV_a1mu0_5_App}{\cb{a}} shows $c/T$ for various values of $h$, down to $T=0.075 \approx 3T_c(h_{c})$. At all $h$ values, $c/T$ is close to that of the tight-binding model (solid curve) for $T\gtrsim 0.15$. The local maximum in the tight-binding $c/T$ is due to the van Hove singularity. At $T<0.15$, $c/T$ is moderately enhanced relative to that of the tight binding model. However, the enhancement occurs both close to and away from the critical point. The overall behavior is qualitatively similar to that of $c/T$ with $\mu=-1$, shown in Fig. 1b of the main text, although some details are different. 
\looseness=-1

The behavior of the nematic susceptibility near the critical point for the different values of $h$ is presented in Fig.~\ref{fig:Inv_chi_App}{\cb{b}}. At $h\approx h_{c}$, the inverse nematic susceptibility goes linearly to zero.
$\tilde{\nu}_0$ for the same parameters is depicted in Fig.~\ref{fig:DOS_a1mu0_5_App}{\cb{c}}, indicating a sharp decrease at $T\approx 3T_c \approx 0.08$. Similarly to the results presented in Fig.~\ref{fig:GF_main}, the superconducting fluctuations seem to onset at a significantly higher temperature than the actual $T_c$. 

\section{Finite size effects} 
\label{Finite_size_Cv }
\begin{figure}
	\centering\includegraphics[width=\columnwidth ]{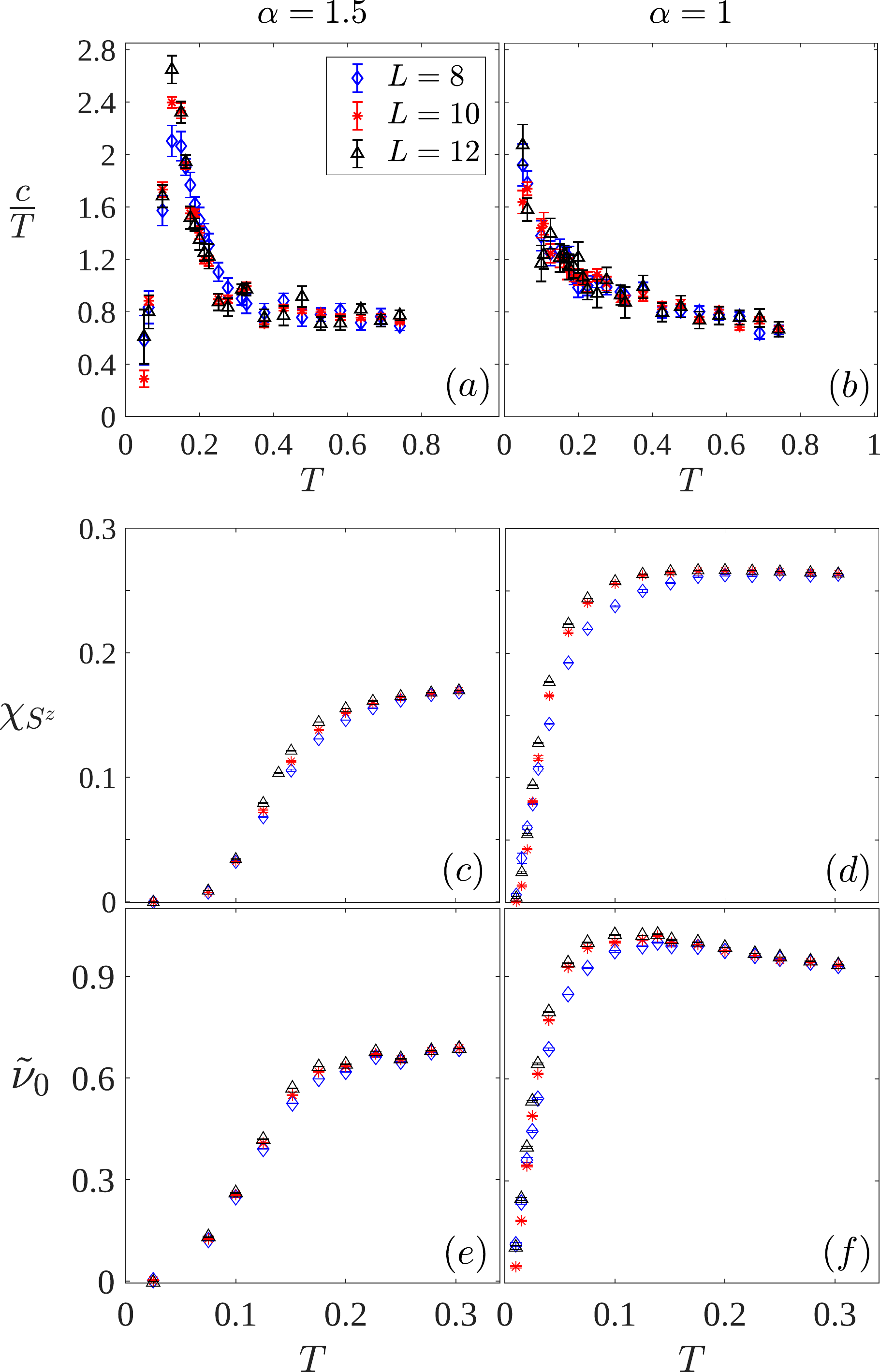}
	\caption{ $c/T$ (a,b), $\chi_{S^z}$ (c,d), and $\tilde{\nu}_0$ (e,f) as a function of temperature for three system sizes $L=8$, $10$, $12$ and two coupling constants $\alpha=1.5$, $1$ at $h \approx  h_c$ and $\mu=-1$. In \textbf{(a,c,e)} $h=2.6$, and in \textbf{(b,d,f)} $h=4.06$.}  
\label{fig:DOS_size_effects}\label{fig:CV_finite_size}\label{fig:chi_sz_finite_size}
\end{figure}

To verify that our conclusions do not depend on the system size, examined the finite size dependence of the results. The nematic susceptibility has been shown in Fig.~2 of the main text for $L=8,L=10$. 
Fig.~\ref{fig:CV_finite_size}{\cb{a,b}} shows $c/T$ for $L=8,10,12$. The spin susceptibility $\chi_{S^z}$ can be found in Fig.~\ref{fig:chi_sz_finite_size}{\cb{c,d}} for the same system sizes. Lastly, we plot $\tilde{\nu}_0 $ in 
Fig.~\ref{fig:DOS_size_effects}{\cb{e,f}}. All of these results are evaluated near criticality ($h\approx h_c$), where the strongest finite-size effects are expected.
We do not observe any substantial finite-size effects, indicating that the system sizes we have simulated are sufficient in order to characterise the thermodynamic limit, at least qualitatively.

\section{Perturbation theory}
\subsection{Specific heat}
We consider the following effective Hamiltonian
\begin{align}
\label{eq:model_RPA}
H_{RPA}&=\sum_{\alpha,\bm{k}}
\epsilon_{\bm{k}}\psi_ {\alpha,\bm{k}}^{\dagger}\psi^{\phantom{\dagger}}_{\alpha,\bm{k}} 
\notag\\&\quad+
\smashoperator{\sum_{\substack{\alpha,\beta\\\bm{q},\bm{k},\bm{k'}}}}
 U(\bm{q},\bm{k},\bm{k'})
 \psi_ {\alpha,\bm{k}-\bm{q}}^{\dagger}
 \psi_ {\beta,\bm{k'}+\bm{q}}^{\dagger}
 \psi_ {\beta,\bm{k'}}^{\phantom{\dagger}}
 \psi_ {\alpha,\bm{k}}^{\phantom{\dagger}}\, ,
\end{align}
where $\varepsilon_{\bm{k}}=(k^2-k_F^2)/2m$, $\alpha,\beta=1,\dots,N$ are fermion flavor indices (the physical case corresponds to $N=2$), and
\begin{align}
U(\bm{q},\bm{k},\bm{k'})&=\frac{g^2f_{\bm{k}-\bm{q}/2}f_{\bm{k'}+\bm{q}/2}}{N(r_0+|\bm{q}|^{2})}.
\end{align}
The system can be tuned to the vicinity of a nematic QCP by tuning $r_0$.
The nematic form factor is $f_{\bm{k}}=\cos k_{x}-\cos k_{y}$. The specific heat follows from the free energy per unit volume $F=\mathcal{F}/L^2$ where $L$ is the linear system size, according to
\begin{align}
    c&=-T\frac{\partial^2 F}{\partial T^2}.
\end{align}
In the following, we employ a canonical field-theoretic formulation of the problem.
In the RPA approximation, justified formally in the large-$N$ limit, the interaction contribution to the free energy is given by the sum of all rings of particle-hole bubbles, i.e.
\begin{align}
F_{\mathrm{RPA}}&=-\frac{T}{2L^2}\sum_{\bm{q},i\omega_m}\sum_{n=1}^{\infty}\frac{1}{n}\biggl[
\frac{g^2}{N(r_0+|\bm{q}|^{2})}
\Pi(\bm{q},i\omega_m)\biggr]^{n}
\label{eq:FRPA}
\shortintertext{with the nematic correlation function $\Pi(\bm{q})$ being defined as}
\Pi(\bm{q},i\omega_m)&=\frac{2T}{L^2}\sum_{k,i\Omega_n}
f_{\bm{k}-\bm{q}/2}f_{\bm{k}+\bm{q}/2}
\notag\\&\quad\times
G_{\bm{k},i\Omega_n}G_{\bm{k}+\bm{q},i\Omega_n+i\omega_m}.
\label{eq:fulllindhard}
\end{align}
For the problem at hand, a number of simplifying assumptions can be taken for the nematic correlation function (Lindhard function). Most importantly, since $T\ll E_F$, to leading order, the temperature dependence of $\Pi$ can be disregarded. Near the QCP, the typical transfer momentum $|\bm{q}|$ is much smaller than the Fermi momentum $|\bm{k}_F|$. Hence, the form factors reduce to $f_{\bm{k}-\bm{q}/2}f_{\bm{k}+\bm{q}/2}\approx f_{\bm{k}}^2$. 
In summary, one recovers the standard low-energy approximation for the density-density correlation function
\begin{align}
\Pi(\bm{q},i\omega_m)
&=-\frac{m f^2_{k_F \bm{\hat q}\times\bm{\hat z}}}{\pi}
\left[1-\biggl( 1+\left(\frac{v_Fq}{\omega_m}\right)^2\biggr)^{-\frac{1}{2}}\right],
\label{eq:lindhard}
\end{align}
As it turns out, the free energy is dominated by regions in phase space where $v_F|\bm{q}|\gg |\omega_m|$, such that the square root in Eq.~\eqref{eq:lindhard} can be replaced by $\frac{|\omega_m|}{v_F|\bm{q}|}$.
Performing the sum over $n$ in Eq.~\eqref{eq:FRPA} to obtain a logarithm and then keeping only the temperature dependent piece of $F_{\mathrm{RPA}}$, we are left with
\begin{align}
F_{\mathrm{RPA}}
&= \frac{T}{4\pi}
\sum_{\omega_m} \int q\mathrm{d} q
\log \left(A q^2	
+\frac{{\left| \omega_m \right|}}{q}   \right),
\label{eq:simplestRPA}
\end{align}
where $A=\frac{\pi  v_F}{2mg^2}$.
The Matsubara summation can be turned into the contour integral over the upper half plane in $z$,
\begin{align}
F_{\mathrm{RPA}}
&= \frac{1}{8\pi^2 i}
\oint_{C} \int_{0}^{\infty } q\mathrm{d} q
\log \left( Aq^2	-\frac{{iz}}{q}   \right) \coth \left(\tfrac{\beta z}{2}\right).
\end{align}
This integral is divergent, but $\partial F_{\mathrm{RPA}}/\partial T$ is finite,
\begin{align}
\frac{\partial F_{\mathrm{RPA}}}{\partial T}
&=-\frac{1}{8\pi^2 T^2}
\int_{0}^{\infty }\!\!\! \mathrm{d} z
\int_{0}^{\infty}\!\!\! q\mathrm{d} q
\frac{z \arctan (\frac{z}{Aq^3}) }{\sinh^2 (\frac{\beta z}{2})}.
\notag\\
&=- \frac{I}{16\pi}\frac{T^{\frac{2}{3}}}{A^{\frac{2}{3}} }
\shortintertext{with}
I &= \int_{0}^{\infty } \mathrm{d} x
\frac{x^{\frac{5}{3}}}{\sinh^2 (\frac{x}{2})} \approx12.78 
\end{align}
Together with the free fermion part, the specific heat is thus given by $c(T)=c_0(T)+c_Q(T)$ with
\begin{align}
  \label{eq:c0}
  c_{0}(T)&=
  \frac{\pi}{6}mNT\\
  c_{Q}(T)&=
  \frac{\pi}{6} J m E_F^{-\frac{1}{3}}g^{\frac{4}{3}}    T^{\frac{2}{3}},
  \label{eq:totalcVRPA}
\end{align}
\begin{figure}
    \centering
    \includegraphics[width=\columnwidth]{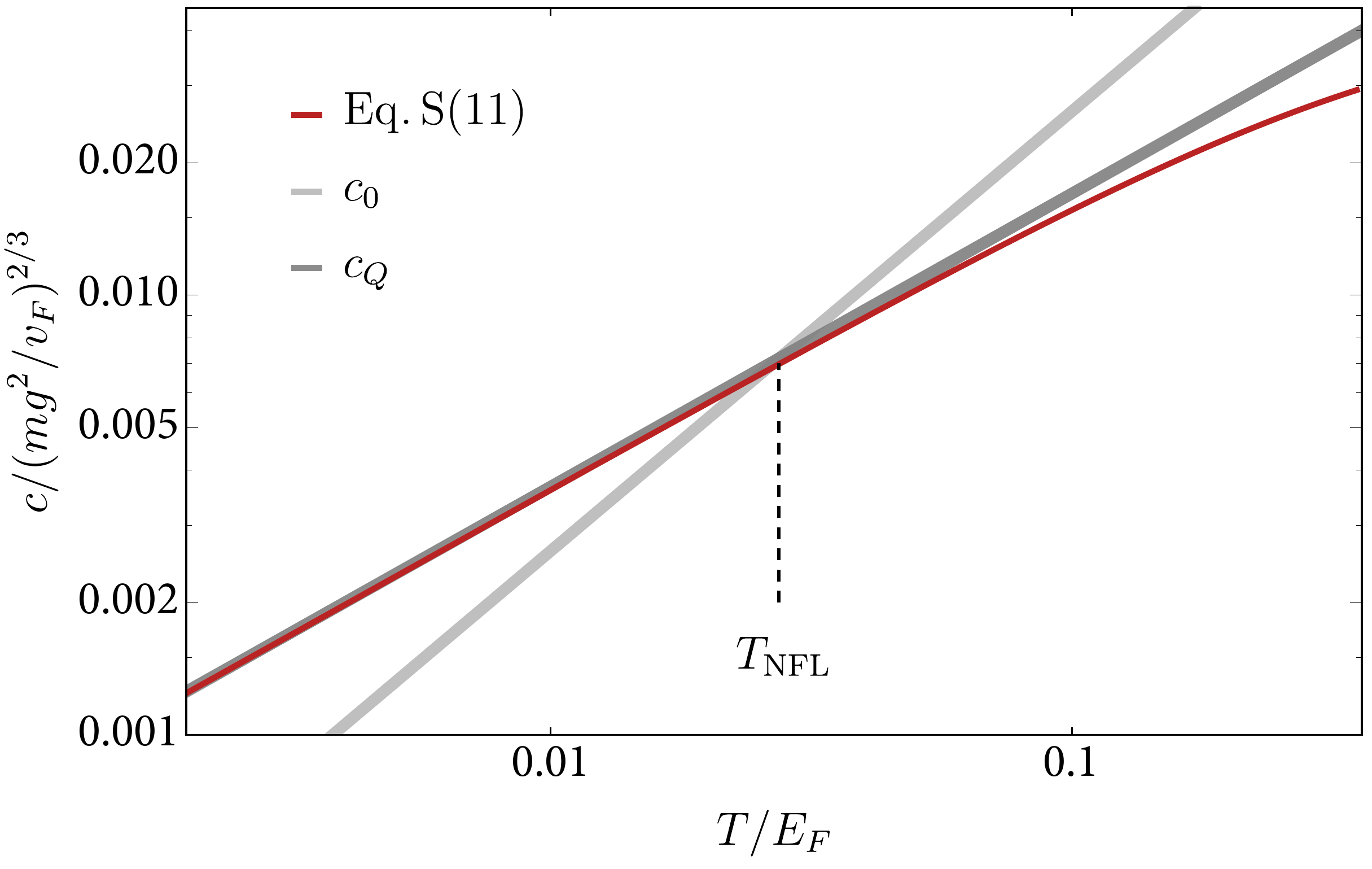}
    \caption{
    Logarithmic plot of the interaction contribution to the specific heat according to Eqs.~(\ref{eq:FRPA},\ref{eq:lindhard}) (red), compared with $c_0$ and $c_Q$ according to the simplified expressions of Eqs.~(\ref{eq:c0},\ref{eq:totalcVRPA}). For the numerical evaluation a high-energy cutoff of $100E_F$ was used. 
    }
    \label{fig:RPA}
\end{figure}
with $J=\frac{I}{2^{5/3}\pi^{8/3}}\approx 0.19$, as mentioned in the main text. 
We also confirmed numerically that the simplifying assumptions taken in this derivation are justified. 
To this end, we compare in Fig.~\ref{fig:RPA} the specific heat of  Eq.~\eqref{eq:totalcVRPA} with the more general form as given by Eqs.~(\ref{eq:FRPA},\ref{eq:lindhard}). 
The crossover temperature $T_{\mathrm{NFL}}$ is given by $c_0(T_{\mathrm{NFL}})=c_Q(T_{\mathrm{NFL}})$, which yields $T_{\mathrm{NFL}}\sim g^4/(N^3 E_F)$, as stated in the main text. At $T_{\mathrm{NFL}}$, the RPA approximation is known to break down, even in the large-$N$ limit~\cite{Lee2009,Metlitski2010,Holder2015}.

\begin{figure}
    \centering
    \includegraphics[width=\columnwidth]{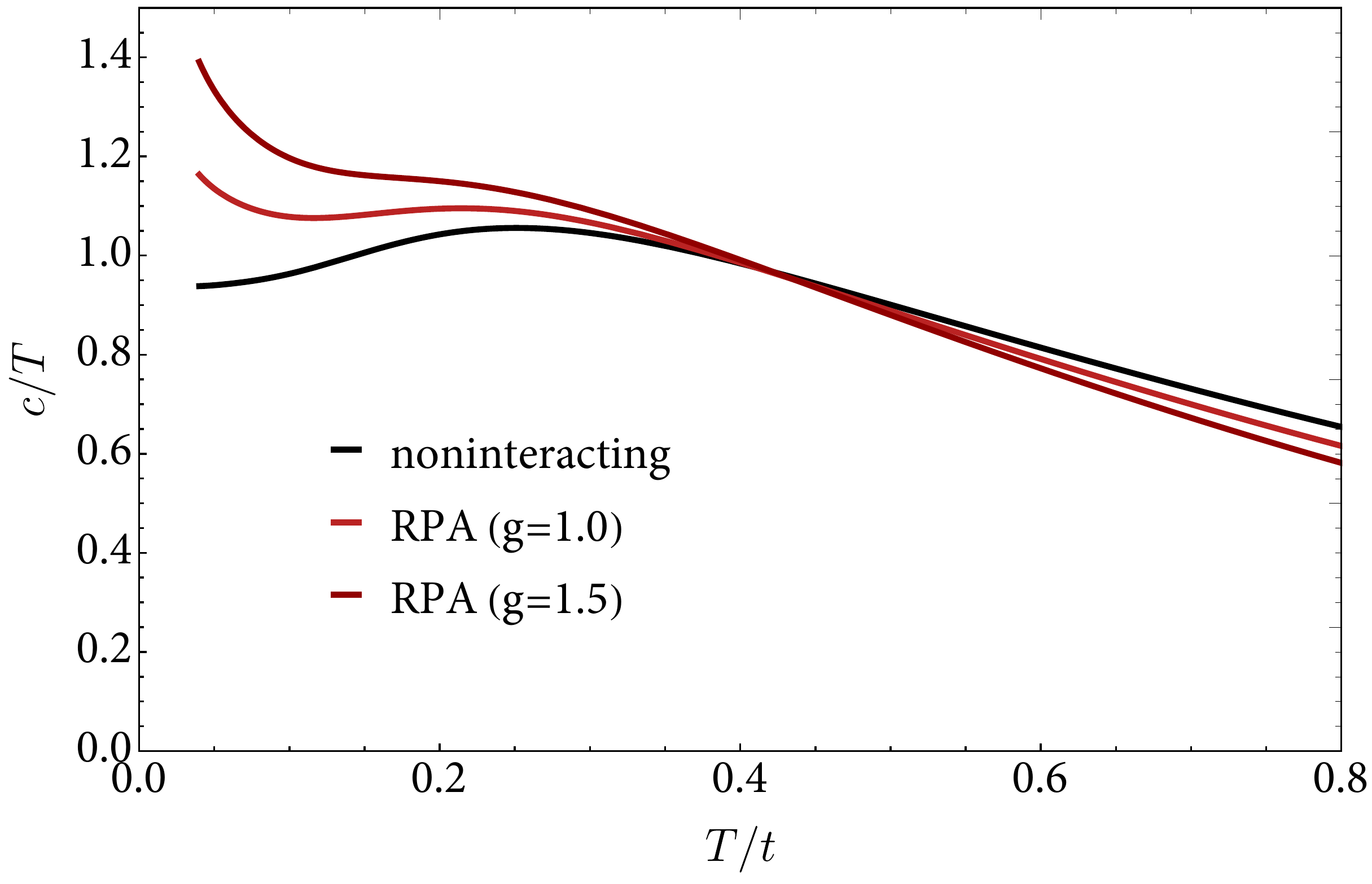}
    \caption{
    $c/T$ as a function of temperature within RPA, using the tight binding dispersion considered in the main text [Eq.~(1)]. Two sets of parameters were used: $g=1.0$, $r_0=0.3 k_F^2T/t$ and $g=1.5$, $r_0=0.4k_F^2T/t$, respectively (the $T$ dependence of $r_0$ was estimated from the QMC results shown in Fig. 2 of the main text). The obtained values for $c_Q/T^{2/3}$ are within a factor of 2 compared to the ones predicted by a free fermion dispersion with the same average effective mass.
    }
    \label{fig:RPA2}
\end{figure}

To account for lattice effects, we have also calculated the specific heat within the RPA using the tight binding dispersion of the main text [Eq.~(1)]. The form factors were approximated as $f_{\bm{k}-\bm{q}/2}f_{\bm{k}+\bm{q}/2}\approx f_{\bm{k}}^2$. The resulting $c/T$ vs. $T$ is shown in Fig.~\ref{fig:RPA2} for two values of the coupling $g$. In agreement with the QMC results, we find that for these values of $g$, $c/T$ remains within $10\%$ of the non-interacting value for $T \gtrsim 0.2t$.

\subsection{Superconducting critical temperature}

We now briefly comment on the expected superconducting transition temperature, $\widetilde{T}_c$,  within the pertubative RPA analysis. 
The scaling of $\widetilde{T}_c$ can be gleaned from the linearized Eliashberg gap equation, that reads~\cite{Wu2020,Marsiglio2008},
\begin{align}
    \Delta(\omega_m)
    &=
    \frac{g^{2}}{Nv_F} \widetilde{T}_c
    \sum_{m'\neq m}
    \frac{\Delta(\omega'_m)-\Delta(\omega_m)\omega'_m/\omega_m}{|\omega'_m|}
    \notag\\&\quad\times
    \int\mathrm{d} q D(q,\omega_m-\omega'_m),
\end{align}
where the dressed boson propagator is given by (neglecting the form factor for simplicity)
\begin{align}
    D(q,\omega)&=(q^2+g^2\nu |\omega|/q)^{-1},
\end{align}
with $\nu$ the DOS at the Fermi level.
The gap equation thus becomes
\begin{align}
    \Delta(\omega_m)
    &=
    \frac{1}{N}\left(\frac{g^{4}}{v_F^2\nu}\right)^{1/3} \widetilde{T}_c
    \sum_{m'\neq m}
    \frac{\Delta(\omega'_m)-\Delta(\omega_m)\omega'_m/\omega_m}{|\omega'_m||\omega_m-\omega'_m|^{1/3}}
    ,
    \label{eq:Delta}
\end{align}
where here and in the following we suppress factors of order $\mathcal{O}(1)$. Eq.~\eqref{eq:Delta} can be recast in dimensionless form after inserting the Matsubara frequencies in the denominator,
\begin{align}
    \Delta(\omega_m)
    &=
    \frac{1}{N}\left(\frac{g^{4}}{E_F \widetilde{T}_c}\right)^{1/3}
    \sum_{m'\neq m}
    \frac{\Delta(\omega'_m)-\Delta(\omega_m)\omega'_m/\omega_m}{|2m'+1||m-m'|^{1/3}}
    .\label{eq:gapeq}
\end{align}
Therefore, the superconducting transition temperature is of order $\widetilde{T}_c\sim g^4/N^3 E_F$, in accordance with similar estimates in the literature~\cite{Wang2016c,Wu2020}. We emphasize that this derivation does not establish whether a superconducting instability occurs at all. This is since the RPA treatment breaks down at $\widetilde{T}_c\sim T_{\text{NFL}}$. The RPA analysis merely indicates that at this scale, the superconducting susceptibility becomes large, and superconducting fluctuations need to be treated on equal footing as the quantum critical fluctuations. 

\subsection{Crossovers in the electron Green's function near a QCP}

In this section, we discuss the general structure expected in the electron self-energy near a QCP. In particular, we emphasize the possible existence of crossover regime where the electronic quasiparticles are coherent and their effective mass enhancement is small, but nevertheless their scattering rate is different from that of a Fermi liquid. 

For concreteness, let us assume that the electron self-energy near the QCP has a power-law form:
\begin{equation}
    \Sigma(i\omega_n) = i (T_{\text{NFL}})^{1-\alpha} \text{sign}(\omega_n) |\omega_n|^{\alpha},
\end{equation}
with $0<\alpha<1$. Analytically continuing to real frequency, we find that both the real and imaginary parts of $\Sigma(\omega)$, $\Sigma'$ and $\Sigma''$ respectively, are proportional to $(T_{\text{NFL}})^{1-\alpha} |\omega|^\alpha$. Thus, in terms of the effective frequency-dependent quasiparticle weight $Z(\omega)=(1 - \frac{\partial \Sigma'}{\partial \omega})^{-1}$, we can define a frequency-dependent effective mass enhancement: $\frac{m^*(\omega)}{m} = [Z(\omega)]^{-1}$, where $m$ is the bare mass. The effective mass enhancement becomes $O(1)$ at $\omega \sim T_{\text{NFL}}$. 
If $T_{\text{NFL}}$ is much smaller than the Fermi energy $E_F$, there is a broad frequency regime where $T_{\text{NFL}} \ll \omega \ll E_F$ where the effective mass enhancement is small. Nevertheless, in the same frequency regime, the quasiparticle scattering rate, $\Gamma(\omega) = Z(\omega) \Sigma''(\omega) \sim \omega^{\alpha}$, has a markedly non-Fermi liquid behavior. Note also that in this regime, $\Gamma(
\omega)\ll \omega$; thus, the quasiparticles are still long-lived, despite having a non-Fermi liquid dependence of the scattering rate on energy. 

The condition that $T_{\text{NFL}}$ is much smaller than $E_F$ is naturally satisfied if the interactions are weak. Surprisingly, our simulations indicate that $T_{\text{NFL}}$ remains substantially smaller than $E_F$ even for moderate to strong interactions, as can be seen by the small enhancement of $c/T$ compared to the band structure value (Fig. 1 of the main text).

\end{document}